\documentclass{iau} 
\usepackage{graphicx}
\usepackage{sidecap}
\usepackage{wrapfig}

\title[The age-kinematical features] 
{The age-kinematical features in the Milky Way outer disk}

\author[Chao Liu, Hai-Jun Tian \& Jun-Chen Wan]   
{Chao Liu$^1$,
 Hai-Jun Tian$^2$ 
 \and Jun-Chen Wan$^{1,3}$}

\affiliation{$^1$Key Lab of Astronomy, National Astronomical Observatories, CAS, Beijing 100012, China \\ email: {\tt liuchao@nao.cas.cn} \\[\affilskip]
$^2$China Three Gorges University, Yichang 443002, China\\[\affilskip]
$^3$ University of China Academy of Sciences, Beijing 100049, China}

\pubyear{2016}
\volume{321}  
\jname{Formation and Evolution of Galaxy Outskirts}
\editors{A. Gil de Paz, eds.}
\begin{document}
\maketitle
\begin{abstract}
We derive the mean velocity components at various Galactocentric radii from 8 to 14 kpc using about 40,000 red clump stars observed in the LAMOST survey. We find that the vertical bulk motion for younger red clump stars are significantly larger than that for the older red clump stars. This is likely the kinematical feature of the Galactic warp around its line-of-node, which is located close to the Galactic anti-center region. It is evident that the warp are mainly contributed by the younger stars rather than the older stars. The age variation in the vertical kinematics favors a formation scenario where the Galactic warp is originated from infalling misaligned gas. 
\end{abstract}

\firstsection 
\section{Introduction}
The outskirt of a disk galaxy can reflect how the disk secularly evolves and therefore is crucial in the studies of the galaxies. This is particularly important in the case of the Milky Way, because we can easily observe the motion and chemical abundance for individual stars, which can be used for understanding the secular evolution of galactic disks. HI surveys have revealed that the Galactic gaseous disk is warped in its outskirts~(\cite[Levine et al. 2006]{levine06}). In the 2nd quadrant of the disk, the HI disk tends to bend up toward north and in the 3rd quadrant, it tends to bend down toward south. The stellar disk also has a similar warp~(\cite[Lopez-Corredoira et al. 2002]{lopez02}). According to the simple kinematics, if the vertical bending amplitude of bending of the warp reaches its maximum in the 2nd and 3rd quadrants, it should reach the maximum vertical velocity in between these two quadrants~(\cite[Lopez-Corredoira et al. 2014]{lopez14}), namely, along the line-of-nodes direction (for the definition of the line-of-node see~\cite[Ro\v{s}kar et al. 2010]{roskar10}). For the case of our galaxy, the line-of-nodes direction is around the Galactic anti-center.

The LAMOST survey has been collecting more than 4 million low resolution stellar spectra in its DR2 catalog~(\cite[Cui et al. 2012]{cui12}, \cite[Zhao et al. 2012]{zhao12}, \cite[Deng et al. 2012]{deng12}). Because the special condition of the location of the telescope, it can well observe the Galactic anti-center region, making the survey one of the best to study the outer disk structures, e.g. the Galactic warp, using the kinematics of a large amount of stars (\cite[Yao et al. 2012]{yao12}, \cite[Liu et al. 2014]{liuxw14}). In this talk, we demonstrate that the selected LAMOST data can well address the age-kinematical features which may be associated with the Galactic warp.

\section{Data and Method}
Recently, \cite[Wan et al. (2015)]{wan15} identified about 120 thousands red clump (RC) candidates from the LAMOST DR2 survey and gave distance estimates with uncertainty of about 10\%. \cite[Tian et al. (2016)]{tian16} further purify the sample and separate them into young ($\lesssim2$\,Gyr) and old ($\gtrsim2$\,Gyr) populations based on the direct comparison with synthetic stellar isochrones. Within $|z|<1$ kpc and $8<R<14$ kpc, they obtained about 40000 RC stars, around one third of which are young.

The LAMOST survey only provides line-of-sight velocities for these distant stars without reliable proper motions. Therefore, we derive the mean radial ($\langle v_R\rangle$), azimuthal ($\langle v_\phi\rangle$), and vertical ($\langle v_z\rangle$) velocity under the Galactocentric cylindrical coordinates for a given $R$ bin according to the following projection relation:
\begin{equation}
v_{los}=-\langle v_R\rangle\cos{(l+\phi)}\cos{b}+\langle v_\phi\rangle \sin{(l+\phi)}\cos{b}+\langle v_z\rangle\sin{b},
\end{equation}
where, $v_{los}$ is the measured line-of-sight velocity, and $l$, $b$, and $\phi$ are the Galactic longitude, latitude, and azimuth angle w.r.t. the Galactic center, respectively. The Monte Carlo simulation confirms that for the LAMOST RC stars within $|z|<1$\,kpc, the uncertainty of $\langle v_z\rangle$ is only about 5\,kms$^{-1}$ at $R<11$\,kpc and becomes $\sim10$\,kms$^{-1}$ at $R\sim13$\,kpc.
\begin{wrapfigure}{r}{0.5\textwidth}
\centering
\includegraphics[scale=0.3]{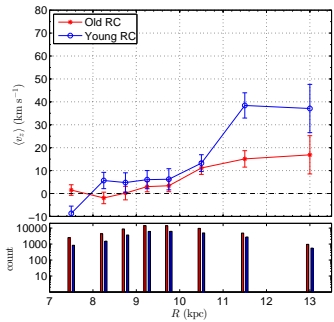}
\caption{The top panel shows $\langle v_z\rangle$ as a function of $R$ for the young (blue circles) and old (red asterisks) RC stars. The bottom panel shows the star number for the young (blue) and old (red) RC stars in each $R$ bin.  All the bins contain more than 500 RC stars to ensure proper statistics.\label{fig1}}
\end{wrapfigure}

\section{Results and conclusions}
Fig.~1 shows  $\langle v_z \rangle$ as a function of $R$ for old (red) and young (blue) RC samples. It is clearly seen that the vertical bulk motion of the young RC stars is significantly larger than that of the old stars. Particularly at $R>11$\,kpc, the mean vertical velocity for the young RC stars is almost 40\,km$s^{-1}$, while for the old RC stars the value is less than 20\,kms$^{-1}$.

To further investigate if the net upward vertical motion is due to the oscillation of the disk discovered by \cite[Carlin et al. (2013)]{carlin13} and \cite[Williams et al. (2013)]{williams13}, we separate them into two sub-samples, one covering $-1.0<z<-0.3$\,kpc and other covering $0.3<z<1.0$\,kpc. The left panel of Fig.~2 shows $\langle v_z\rangle ( R)$ for the southern sub-sample and the right panel for the northern. It is found that the young RC stars show positive $\langle v_z\rangle$ in both northern and southern sub-samples when $R>11$\,kpc, while the old RC stars show short wave (along $R$) oscillations. 

For the old RC stars, at $8<R<10.5$\,kpc, they are in a compression process, i.e. the stars located above (below) the mid-plane averagely move downward (upward), which is in consistent with \cite[Carlin et al. (2013)]{carlin13}. However, at $R\sim11.5$\,kpc, they seem to be in an expansion process. These oscillations are moderate in amplitude with $|\langle v_z\rangle|\lesssim10$\,kms$^{-1}$. Although the bulk motion of the young stars similarly oscillate within $R<11$\,kpc, they together move upward when $R>11$\,kpc with a large amplitude of 30--40\,kms$^{-1}$. This is very hard to be explain by the perturbation of a merging event (\cite[G\'omez et al. 2013]{gomez13}) or the spiral arm (\cite[Debattista 2014]{debattista2014}) since the vertical velocity is too large for the perturbation. Compared to the model given by \cite[L\'opez-Corredoira et al. (2014)]{lopez14}, the large vertical bulk motion for the young population is very likely due to the Galactic warp. Because the Galactic anti-center region, which is the one sampled by the LAMOST survey, is approximately the same direction of the line-of-nodes of the warp, the mean vertical velocity of the stars in the warp is expected to be around its maximum. This can explain the large vertical bulk motion for the young RC stars. However, the old RC stars only show very weak upward bulk motion, implying that it may not belong to the warp in which the young RC stars are embedded.
\begin{figure}[bth]
\centering
\begin{minipage}{15cm}
\centering
\includegraphics[scale=0.3]{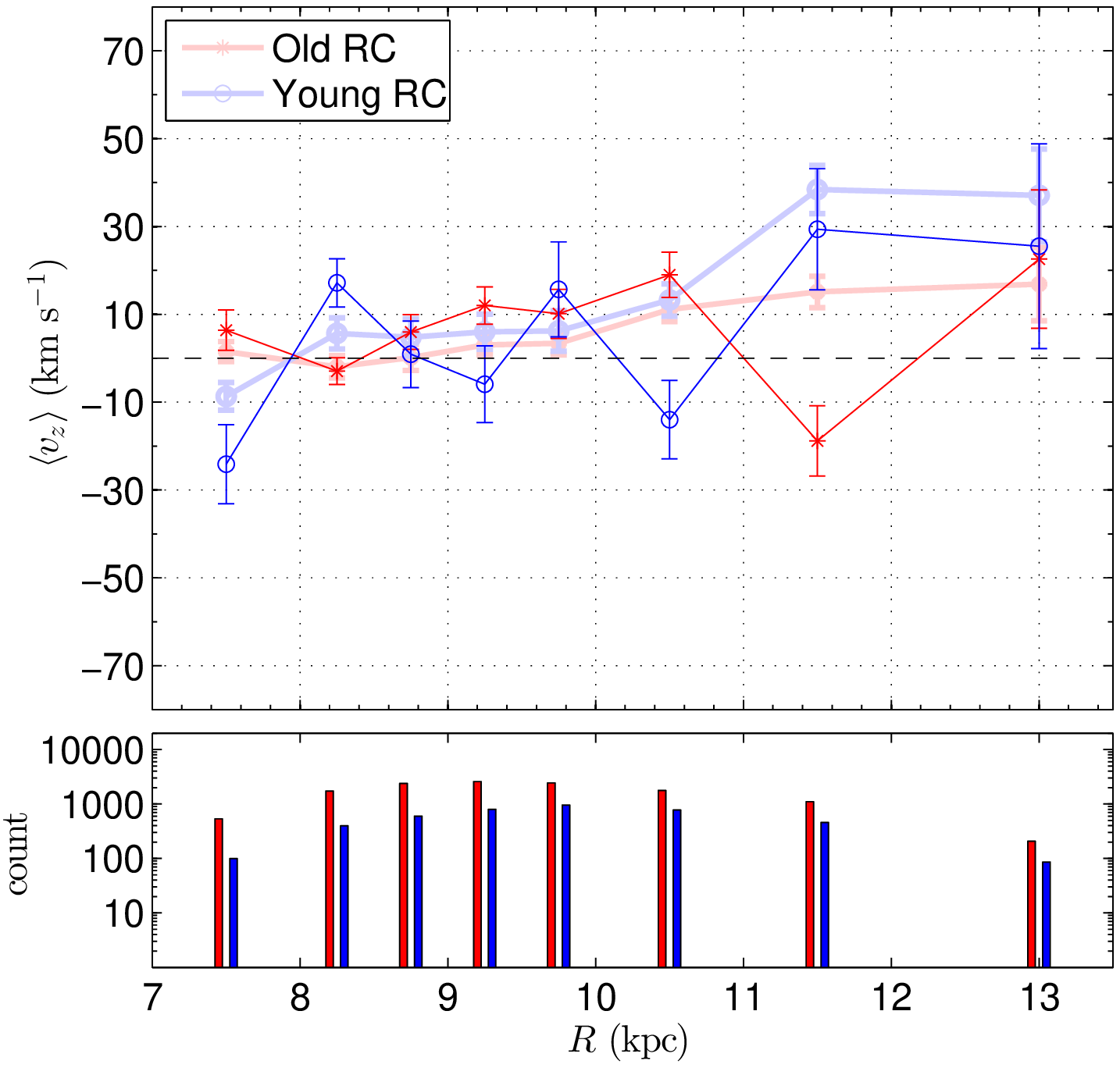}
\includegraphics[scale=0.3]{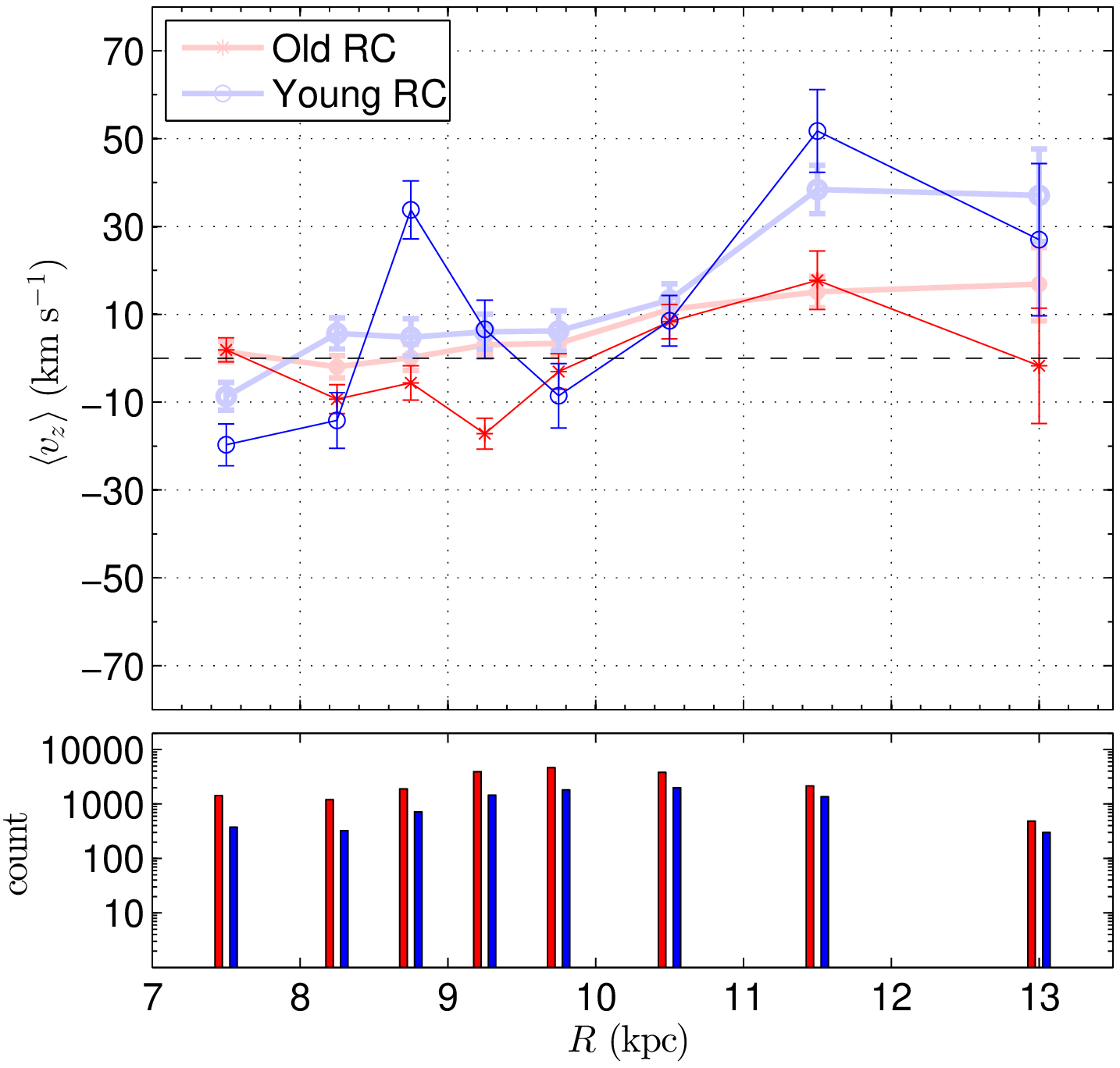}
\end{minipage}
\caption{The left panel shows $\langle v_z\rangle(R)$ for the young (blue circles) and old (red asterisks) RC populations with $-1<z<-0.3$\,kpc. As a comparison, the piled blue and red lines indicate the location of $\langle v_z\rangle$ for the full sample of the young and old populations, respectively. The right panel is similar but for the data with $+0.3<z<+1.0$\,kpc.\label{fig2}}
\end{figure}

In summary, the new LAMOST survey data reveals the age-related mean vertical motions beyond $R=11$\,kpc for the RC samples. It is evident that different ages of the stars may belong to the warps with different vertical bending amplitude. Such an age variation in the warp challenges the theoretical scenario that the Galactic warp is formed from dynamical torque (\cite[Shen et al. 2006]{shen06}) but favors the scenario that the warp is formed from infalling misaligned gas, in which star formation can be triggered when the gas cools down (\cite[Ro\v{s}kar et al. 2010]{roskar10}). 


\end{document}